\documentclass[aps,showpacs,preprintnumbers,amsmath, amssymb]{revtex4}

\usepackage{float}
\usepackage{graphics,epsfig}
\usepackage{graphicx}
\usepackage{dcolumn}
\usepackage{bm}
\usepackage{amsmath}
\usepackage{mathrsfs}

\begin{document}
\baselineskip=0.8 cm

\title{{\bf Investigations on hoop conjecture for horizonless spherical charged stars}}
\author{Yan Peng$^{1,2}$\footnote{yanpengphy@163.com}}
\affiliation{\\$^{1}$ School of Mathematical Sciences, Qufu Normal University, Qufu, Shandong 273165, China}
\affiliation{\\$^{2}$ Center for Gravitation and Cosmology, College of Physical Science
and Technology, Yangzhou University, Yangzhou 225009, China}

\vspace*{0.2cm}
\begin{abstract}
\baselineskip=0.6 cm
\begin{center}
{\bf Abstract}
\end{center}

For horizonless spherical stars with uniform charge density, the hoop conjecture was
tested based on the interior solution. In this work, we are interested
in more general horizonless spherical charged stars.
We test hoop conjecture using the exterior solution
since all types of interior solutions
correspond to the same exterior Reissner-Nordsr$\ddot{o}$m solution.
Our analysis shows that the hoop conjecture is violated
for very compact stars if we express the conjecture with the total ADM mass.
And the hoop conjecture holds if we express the conjecture using the mass in the sphere.

\end{abstract}

\pacs{11.25.Tq, 04.70.Bw, 74.20.-z}\maketitle
\newpage
\vspace*{0.2cm}

\section{Introduction}

The famous hoop conjecture introduced almost five decades ago
asserts that the existence of black hole horizons is characterized
by the mass and circumference relation $\frac{C}{4\pi \mathcal{M}}\leqslant 1$ \cite{hc1,hc2}.
Here C is the circumference of the smallest ring that can engulf the
black hole in all azimuthal directions and $\mathcal{M}$ is usually interpreted
as the asymptotically measured total ADM mass \cite{hc3}-\cite{ahc9}.

For horizonless curved spacetimes, the hoop conjecture
should be characterized by the opposite inequality $\frac{C}{4\pi \mathcal{M}}> 1$ \cite{hc1,hc2}.
Then it is a question whether the mass in the ratio can still be
interpreted as the ADM mass. For a specific interior solution
of horizonless charged star with uniform charge density,
However, if $\mathcal{M}$ is interpreted as the ADM mass,
the relation $\frac{C}{4\pi \mathcal{M}}> 1$ can be violated
for certain set of parameters and this relation holds if $\mathcal{M}$
is the mass contained in the engulfing sphere \cite{hc19,hc20,hc21,hc22}.
In contrast, black holes can violate the hoop relation
if the mass term is interpreted as the mass in the sphere \cite{hc23}.
Considering the different appearances of hoop conjecture
in black holes and horizonless stars with uniform charge density,
it is interesting to test hoop conjecture in the
background of more general horizonless compact stars.
In particular, it is meaningful to examine the case of
horizonless stars compact nearly to form horizons.

We extend the discussion in \cite{hc19,hc20,hc21,hc22}
by considering the exterior solution since all types of interior
solutions correspond to the same exterior Reissner-Nordsr$\ddot{o}$m solution. For general
horizonless spherical charged stars, our analysis shows that the hoop relation is
violated if the mass term is interpreted as the ADM mass
and the hoop relation holds if we use the gravitating mass
within the sphere.

\section{Studies of the mass in hoop conjecture}

We are interested in general horizonless spherical charged stars.
And the spacetime reads \cite{hc19,hc20,hc21,hc22}
\begin{eqnarray}\label{AdSBH}
ds^{2}&=&-e^{\nu}dt^{2}+e^{\lambda}dr^{2}+r^{2}(d\theta^2+sin^{2}\theta d\phi^{2}).
\end{eqnarray}
The metric functions $\nu$ and $\lambda$ only depend on the radial coordinate r.
The sphere surface radius is located at $r_{0}$.
In the exterior region $r\geqslant r_{0}$, the background is the
Reissner-Nordsr$\ddot{o}$m solution
\begin{eqnarray}\label{AdSBH}
e^{\nu}=e^{-\lambda}=1-\frac{2M}{r}+\frac{Q^2}{r^2},
\end{eqnarray}
where $M$ is the ADM mass of the spacetime and Q is the
star charge. In this work, we pay attention to the case of
$M\geqslant Q$. The would be horizon position is at $r_{h}=M+\sqrt{M^2-Q^2}$.
Since we are interested in horizonless
stars, there is the relation $r_{0}>r_{h}$.
We can simply set the surface radius as $r_{0}=(1+\varepsilon)r_{h}$,
where $\varepsilon$ is a small positive parameter.

The circumference C is given by
\begin{eqnarray}\label{AdSBH}
C=2\pi r_{0}=2\pi (1+\varepsilon)(M+\sqrt{M^2-Q^2}).
\end{eqnarray}

For horizonless curved spacetimes, the hoop conjecture
should be characterized by \cite{hc1,hc2}
\begin{eqnarray}\label{AdSBH}
\frac{C}{4\pi \mathcal{M}}> 1.
\end{eqnarray}

If we interpret the mass $\mathcal{M}$ as the ADM mass M.
The mass to circumference ratio is
\begin{eqnarray}\label{AdSBH}
\frac{C}{4\pi \mathcal{M}}=\frac{2\pi (1+\varepsilon)(M+\sqrt{M^2-Q^2})}{4\pi M}=\frac{(1+\varepsilon)(M+\sqrt{M^2-Q^2})}{2M}.
\end{eqnarray}

For parameters satisfying $\varepsilon\leqslant\frac{M-\sqrt{M^2-Q^2}}{M+\sqrt{M^2-Q^2}}$,
the relation (5) yields that

\begin{eqnarray}\label{AdSBH}
\frac{C}{4\pi \mathcal{M}}\leqslant1.
\end{eqnarray}
So for very compact stars with $\varepsilon\leqslant\frac{M-\sqrt{M^2-Q^2}}{M+\sqrt{M^2-Q^2}}$,
the hoop conjecture is violated if $\mathcal{M}$ is the ADM mass.

If we interpret $\mathcal{M}$ as the mass in the engulfing sphere, the mass is

\begin{eqnarray}\label{AdSBH}
\mathcal{M}=M-\frac{Q^2}{2r_{0}}.
\end{eqnarray}

The hoop conjecture is expressed by the mass to circumference ratio
\begin{eqnarray}\label{AdSBH}
\frac{C}{4\pi \mathcal{M}}=\frac{2\pi r_{0}}{4\pi (M-\frac{Q^2}{2r_{0}})}=\frac{(1+\varepsilon)^2r_{h}^2}{2M(1+\varepsilon)r_{h}-Q^2}=
\frac{r_{h}^2+2\varepsilon r_{h}^2+\varepsilon^2r_{h}^2}{2Mr_{h}-Q^2+2M\varepsilon r_{h}}>
\frac{r_{h}^2+2\varepsilon r_{h}^2}{2Mr_{h}-Q^2+2M\varepsilon r_{h}}.
\end{eqnarray}

Since $r_{h}$ is the horizon satisfying $1-\frac{2M}{r_{h}}+\frac{Q^2}{r_{h}^2}=0$, there is the relation
\begin{eqnarray}\label{AdSBH}
r_{h}^2=2M r_{h}-Q^2.
\end{eqnarray}

Considering $r_{h}=M+\sqrt{M^2-Q^2}\geqslant M$, we get
\begin{eqnarray}\label{AdSBH}
2\varepsilon r_{h}^2\geqslant 2\varepsilon M r_{h}.
\end{eqnarray}

The relations (8), (9) and (10) imply that
\begin{eqnarray}\label{AdSBH}
\frac{C}{4\pi \mathcal{M}}>1.
\end{eqnarray}
It means the hoop conjecture holds for horizonless stars if we
use the mass within the engulfing sphere. Since we consider the exterior solution,
our conclusion holds in the exterior region of various horizonless stars.
Here we analytically show that for exterior solutions of compact stars,
Thorne hoop conjecture may generally hold if the mass term is
interpreted as mass contained within the engulfing sphere.

\section{Conclusions}

The famous hoop conjecture is expressed by the mass to circumference ratio.
For horizonless spherical stars with uniform charge density, the hoop conjecture was
tested based on the interior solution in \cite{hc19,hc20,hc21,hc22}. We investigated hoop conjecture in
the exterior region of more general spherical horizonless compact stars.
We tested hoop conjecture using the exterior solution
since all types of interior solutions correspond to the same
exterior Reissner-Nordsr$\ddot{o}$m solution.
Our analysis showed that the hoop conjecture cannot hold for very compact stars if
the mass is interpreted as the ADM mass in the total spacetime.
And the hoop conjecture holds if we interpret the mass as the gravitating mass contained within
the engulfing sphere.

\begin{acknowledgments}

This work was supported by the Shandong Provincial Natural Science Foundation of China under Grant
No. ZR2018QA008. This work was also supported by a grant from Qufu Normal University
of China under Grant No. xkjjc201906.

\end{acknowledgments}

\end{document}